\documentclass[aps, onecolumn, superscriptaddress]{revtex4}
\usepackage{float}
\usepackage{dcolumn}
\usepackage{amsmath}
\input{epsf}
\input{epsfx}

\ifx\epstexversion\undefined
  \usepackage{graphicx}
\else
  \usepackage[epstex]{graphicx}
\fi

\def\braket#1{\mathinner{\langle{#1}\rangle}}

\newcommand{\fr}[2]{\frac{{#1}}{{#2}}}

\newcommand{\unit}[1]{\hspace{0pt} {#1}}

\begin{document}
\title{Controlling the Spontaneous Emission Rate of Single Quantum Dots in a 2D Photonic Crystal}

\author{Dirk Englund}
\affiliation{Ginzton Laboratory, Stanford University, Stanford CA 94305}
\author{David Fattal}
\affiliation{Ginzton Laboratory, Stanford University, Stanford CA 94305}
\author{Edo Waks}
\affiliation{Ginzton Laboratory, Stanford University, Stanford CA 94305}
\author{Glenn Solomon}
\affiliation{Ginzton Laboratory, Stanford University, Stanford CA 94305}
\affiliation{Solid-State Photonics Laboratory, Stanford University, Stanford CA 94305}
\author{Bingyang Zhang}
\affiliation{Ginzton Laboratory, Stanford University, Stanford CA 94305}
\author{Toshihiro Nakaoka}
\affiliation{Institute of Industrial Science, University of Tokyo, Tokyo, Japan}
\author{Yasuhiko Arakawa}
\affiliation{Institute of Industrial Science, University of Tokyo, Tokyo, Japan}
\author{Yoshihisa Yamamoto}
\affiliation{Ginzton Laboratory, Stanford University, Stanford CA 94305}
\author{Jelena Vu\v{c}kovi\'{c}}
\affiliation{Ginzton Laboratory, Stanford University, Stanford CA 94305}





\date{January 14, 2005}

\begin{abstract}
We observe large spontaneous emission rate modification of individual InAs Quantum Dots (QDs) in 2D a photonic crystal with a modified, high-$Q$ single defect cavity.  Compared to QDs in bulk semiconductor, QDs that are resonant with the cavity show an emission rate increase by up to a factor of 8.  In contrast, off-resonant QDs indicate up to five-fold rate quenching as the local density of optical states (LDOS) is diminished in the photonic crystal.  In both cases we demonstrate photon antibunching, showing that the structure represents an on-demand single photon source with pulse duration from 210\unit{ps} to 8\unit{ns}.  We explain the suppression of QD emission rate using Finite Difference Time Domain (FDTD) simulations and find good agreement with experiment.  

\end{abstract}
\pacs{42.50.Ct, 42.70.Qs, 42.50.Dv, 78.67.Hc}

\maketitle

One of the core issues of modern optics is the subject of photon interaction with matter. In the Wigner-Weisskopf approximation, the emission rate is directly proportional to the LDOS\cite{Scully1997}.  Over the past decade, photonic resonators with increased LDOS have been exploited to enhance emission rate for improving numerous quantum optical devices (e.g., \cite{santori2002, McKeever2003}).   Single photon sources in particular promise to see large improvements\cite{Kiraz2004}.  While more attention has been given to \textit{increasing} emission rate, the reverse is also possible in an environment with decreased LDOS.

Here we demonstrate that by designing a photonic crystal structure with a modified single-defect cavity, we can significantly increase or decrease the spontaneous emission (SE) rate of embedded QDs.  Photonic crystals (PCs), periodic arrays of alternating refractive index, are near-ideal testbeds for such experiments.  Their electromagnetic band structure modifies the LDOS relative to free space and hence the SE rate of embedded QD emitters.  We demonstrate that SE of cavity-coupled QDs is enhanced up to 8 times compared to QDs in bulk GaAs.  This coupling paves the way to single photon sources with higher out-coupling efficiency and visibility.  On the other hand, \textit{decoupled} QDs emit at up to five-fold decreased rate compared to bulk.  This lifetime enhancement is significantly higher than previous reports for photonic crystals\cite{Lodahl2004}.  Moreover, our experiments address individual quantum dots.  

The total SE rate of a QD at position $\vec{r}_A$, spectrally detuned from the cavity resonance wavelength by $\lambda-\lambda_{cav}$, can be expressed as the sum of rates into cavity modes and all other modes, $\Gamma = \Gamma_{cav}+\Gamma_{PC}$.  In the weak-coupling regime where the cavity decay rate $\kappa = \fr{\pi c}{\lambda Q}$ exceeds the QD-cavity coupling strength $|g_{cav}(\vec{r}_A)|$, the SE rates can be calculated from Fermi's Golden Rule.  The cavity density of states follows a Lorentzian and gives a simple expression for $\Gamma_{cav}$\cite{Scully1997}.  Similarly, $\Gamma_{PC}$ is related to the LDOS, which in the PC band gap is reduced relative to bulk semiconductor.  

Comparing $\Gamma$ to the bulk semiconductor emission rate $\Gamma_0$ gives the SE rate enhancement factor, 
\begin{eqnarray}
\label{eq:purcell_factor}
\fr{\Gamma}{\Gamma_0} & = & F_{cav} \left(\fr{\vec{E}(\vec{r}_A) \cdot \vec{\mu}}{|\vec{E}_{\max}| |\vec{\mu}|} \right)^{2} \fr{1}{1+4 Q^{2} (\fr{\lambda}{\lambda_{cav}}-1)^{2}} + F_{PC}
\end{eqnarray}
In this equation, the factors $ \left(\fr{\vec{E}(\vec{r}_A) \cdot \vec{\mu}}{|\vec{E}_{\max}| |\vec{\mu}|} \right)^{2}$ and $\fr{1}{1+4 Q^{2} (\fr{\lambda}{\lambda_{cav}}-1)^{2}}$ describe the spatial and spectral mismatch of the emitter dipole $\vec{\mu}$ to the cavity field $\vec{E}$, and $F_{PC}\equiv \Gamma_{PC}/\Gamma_0$ is the base SE rate in the photonic crystal without cavity.  For perfect alignment, the rate enhancement is given by $F_{cav}+F_{PC}$, where the cavity Purcell factor 
\begin{equation}
F_{cav} \equiv \fr{3}{4\pi^{2}} \fr{\lambda^{3}}{n^{3}} \fr{Q}{V_{mode}}.  
\end{equation}
The refractive index $n=3.6$ in GaAs and the cavity mode volume $V_{mode} \equiv (\int_V \varepsilon(\vec{r}) |\vec{E}(\vec{r})|^{2} d^{3}\vec{r})/\max(\varepsilon(\vec{r}) |\vec{E}(\vec{r})|^{2})$. 

\begin{figure}[htbp]
    \includegraphics[width=4in]{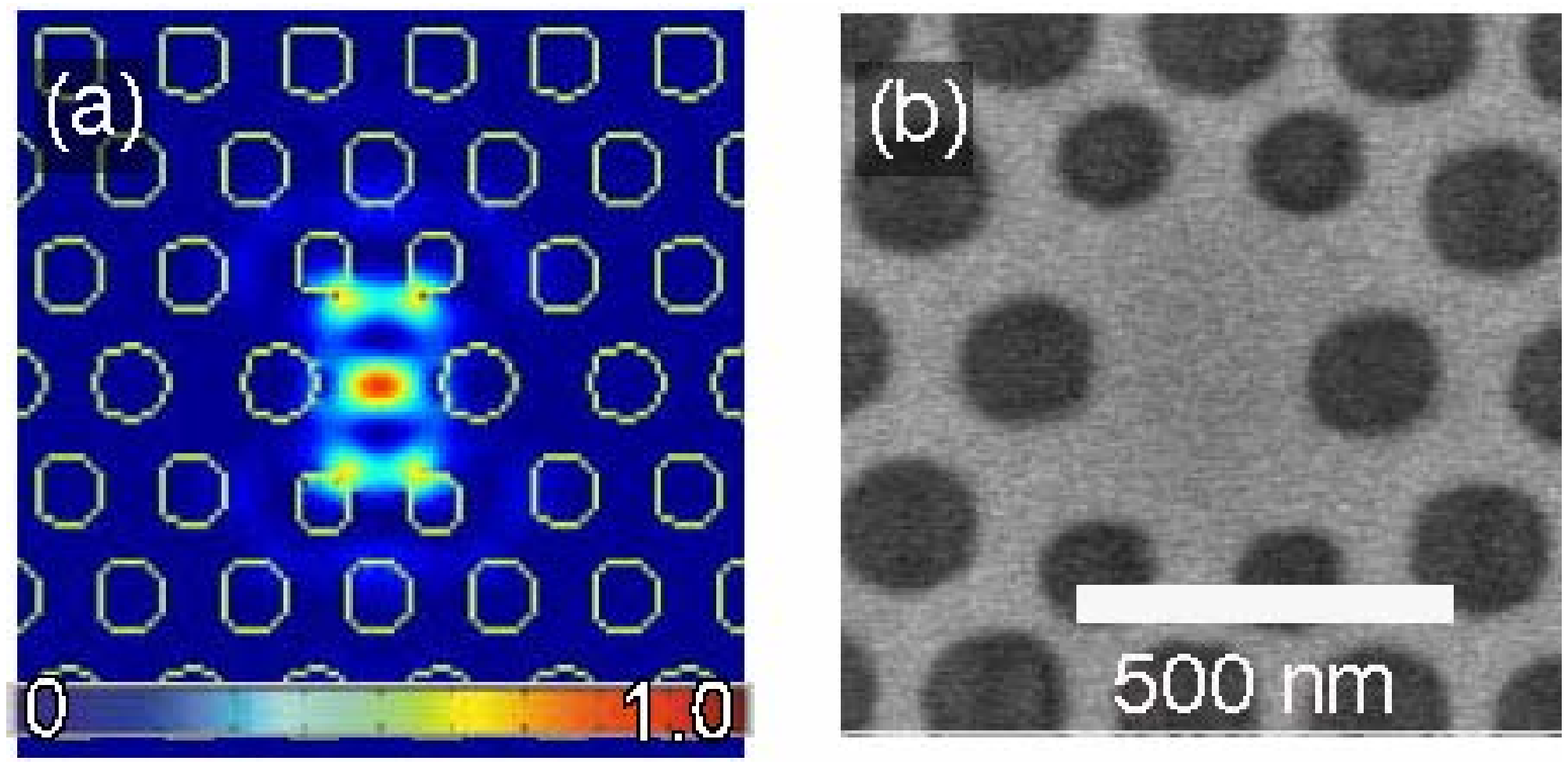}
   \caption{FDTD-assisted design of the photonic crystal cavity.  The periodicity is $a = 0.27 \lambda_{cav}$, hole radius $r=0.3 a$, and thickness $d=0.65 a$.  (a) Electric field intensity of X-dipole resonance.  (b) SEM image of fabricated structure. }
\label{fig:FDTD_cav}
\end{figure}

\begin{figure}[htbp]
    \includegraphics[width=4in]{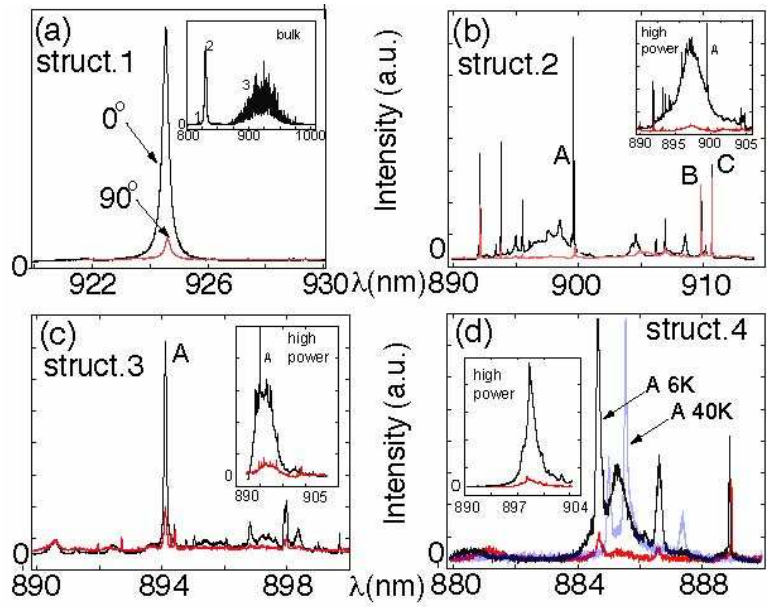}
\caption{PL measurements.  (a) Structure 1 at high pump intensity (1.5\unit{kW/cm$^{2}$}) shows the polarized x-dipole cavity with $Q \approx 5000$.  No single exciton lines match the cavity frequency in this case.  \textit{Inset:} Bulk QD spectrum.  Peak 1: GaAs bandgap transitions; 2: GaAs free and impurity-bound exciton emission; 3: Broadband QD emission.  (b,c): The high pump-power spectra (insets) for structures 2 and 3 reveal resonances with $Q\approx250$ and $Q\approx200$, respectively.  The low-power ($\sim$0.5\unit{kW/cm$^{2}$}) spectra show coupled single exciton lines that match the cavity polarization.  (d) Structure 4 with single exciton line A coupled to dipole cavity mode with $Q\approx 1,600$ (inset).  Temperature-tuning from 6 to 40K allows improved spectral alignment (shaded line).  In all measurements, the Ti-Saph laser was pulsed at 160\unit{fs} with $\lambda=750$\unit{nm}.  } 
\label{fig:PL}
\end{figure}

We designed a single-defect cavity in a 2D-photonic crystal to maximize SE enhancement (Fig.\ref{fig:FDTD_cav}).  The structure was introduced earlier \cite{Vuckovic2003} and modified slightly for fabrication purposes.  Briefly, we use FDTD to design high $Q$, low $V_{mode}$ cavities with maximal QD-field interaction $\braket{\vec{\mu}\cdot \vec{E}(\vec{r})}$.  To better meet fabrication constraints, the design was adjusted based on FDTD analyses of fabricated structures \cite{Englund2004b}.  The cavity supports an x-dipole mode with predicted $Q_{pred} = 45,000$  and volume $\fr{1}{2}(\lambda/n)^{3}$.    We patterned this structure on 160\unit{nm}-thick GaAs membranes by a combination of electron beam lithography, dry- and wet- etching.  Grown by molecular epitaxy, these membranes include a single central layer of self-assembled InAs QDs with density $200/\mu$m$^{2}$.  The QD emission is inhomogeneously distributed about 920\unit{nm} with 50\unit{nm} linewidth (inset Fig.\ref{fig:PL}(a)).  The cavity resonance frequency was chosen to fall near the middle of the QD distribution.

\begin{figure}[htbp]
    \includegraphics[width=4in]{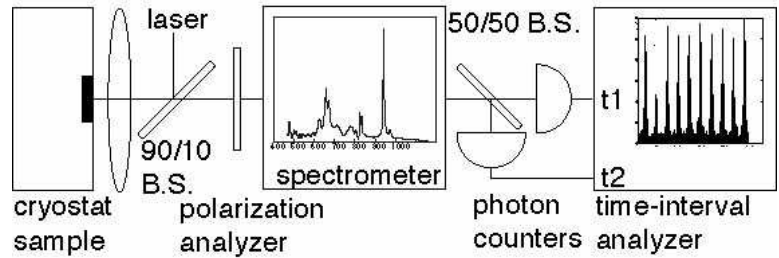}
    \caption{Apparatus for photoluminescence and autocorrelation measurements.
The incident Ti:saph laser beam (continuous or 160\unit{fs}-pulsed) pumps QDs at 5\unit{K}.  The emission is either directly analyzed by the spectrometer (75cm, N$_2$-cooled CCD), or spectrally filtered and sent to the HBT-type setup for coincidence measurements.}
    \label{fig:PL_setup}
\end{figure}

Cavity resonances of fabricated structures were measured by photoluminescence (PL) spectroscopy at 5\unit{K}.  In the confocal microscope setup of Fig.\ref{fig:PL_setup}, the above-bandgap pump beam at 750\unit{nm} excites QDs within the $\sim$600\unit{nm} focal spot.  This spot is much smaller than the 5\unit{$\mu$m}-diameter PC so that only QDs within the crystal are addressed.  The fraction of photoluminescence that originates from the cavity is enhanced over the background in two ways: through enhanced emission rate, and an increased coupling efficiency between cavity mode and collection optics.  The latter efficiency was estimated by FDTD analysis of the cavity mode radiation pattern and equals $\sim 0.07$ for our objective lens with $NA=0.6$.  This enhancement allows us to map out the cavity by pumping the QDs at high intensity, resulting in a broad inhomogeneous emission spectrum that mimics a white light source.  The cavity $Q$ then follows directly from a fit to the Lorentzian in Eq.\ref{eq:purcell_factor}.  The resonance shown in Fig.\ref{fig:PL}(a) is a typical example for a fabricated cavity.  It matches the polarization dependence predicted by FDTD.  The quality factor $Q=5,000$ misses the predicted value $Q_{pred}$ by an order of magnitude.  We believe this decrease to result largely from fabrication inaccuracies as FDTD simulations of fabricated structures, which take these errors into account, do match the measured resonances\cite{Englund2004b}.  We note also that we observed no cavity linewidth narrowing at high pump intensity, so the cold-cavity $Q$ is not over-estimated as a result of stimulated emission. 

These typical high-$Q$ cavities have the disadvantage that spectral coupling is unlikely.  A calculation of the odds for spatial and spectral alignment predicts $\fr{\Gamma}{\Gamma_0} > 20$ in only about 4\% of cavities with $Q \approx 4000$.  To analyze a larger data set, we must therefore focus on the easier coupling to low-$Q$ cavities.  For the present, we will consider three different structures, labeled 2, 3, and 4, with the same cavity design of Fig.\ref{fig:FDTD_cav}.  Figs.\ref{fig:PL}(b)-(d) show their PL spectra.  High pump intensity spectra (insets) indicate x-dipole resonance modes with quality factors $Q_2=200, Q_3=250, Q_4=1600$.   

\begin{figure}[htbp]
    \includegraphics[width=4in]{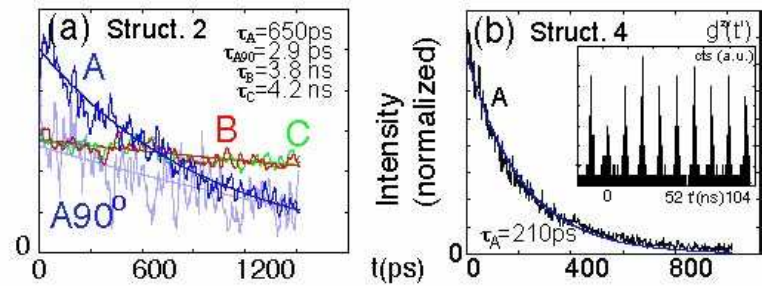}
\caption{Time-resolved measurements of modified single exciton lifetimes.  (a) Structure 2: Lifetime is shortened for the coupled QD line A and extended for the decoupled ones, including the near-degenerate orthogonal emission line A90$^{\circ}$.  (b) Resonant line A of structure 4 shows lifetime shortening to 210\unit{ps}.  The autocorrelation data show antibunching to $g^{(2)}(0) \approx 0.16$ (inset). } 
\label{fig:lifetime}
\end{figure}

In Structure 2, the low-intensity spectrum reveals a single QD exciton line A, spectrally matched with the cavity (Fig.\ref{fig:PL}(b)).  Because of the low QD-cavity coupling probability, further evidence is required to confirm QD-cavity coupling.  This comes from polarization matching.  If the dot is coupled, only one of its two near-degenerate orthogonal emission lines \cite{Hoegele2004} is enhanced.  Fig.\ref{fig:PL}(b) indicates this is the case.  For the coupled system, we expect the excitonic lifetime $\tau\equiv\fr{1}{\Gamma}$ to be diminished by the Purcell factor compared to the surrounding.  Using a streak camera with temporal resolution of 50\unit{ps}, we measured $\tau_A = 650$\unit{ps} (Fig.\ref{fig:lifetime}(a)).  Compared to the excitonic lifetime in bulk semiconductor, which has a distribution of $\tau_0 = 1.7\pm0.3$\unit{ns}, this indicates a rate enhancement by $F_A=2.6\pm0.5$.  In striking contrast is the lifetime of QD excitons \textit{not} coupled to a cavity.  The cross-polarized emission A90$^{\circ}$ has a lifetime of 2.9\unit{ns}, far longer than in bulk.  As can be seen in Fig.\ref{fig:PL}(b), this line is much weaker than A due to low outcoupling and SE rate, as expected.  Lines B and C, which are spectrally far detuned, show lifetimes extended even further to $3.8$\unit{ns} and $4.2$\unit{ns}, respectively.  

\begin{figure}[htbp]
    \includegraphics[width=4in]{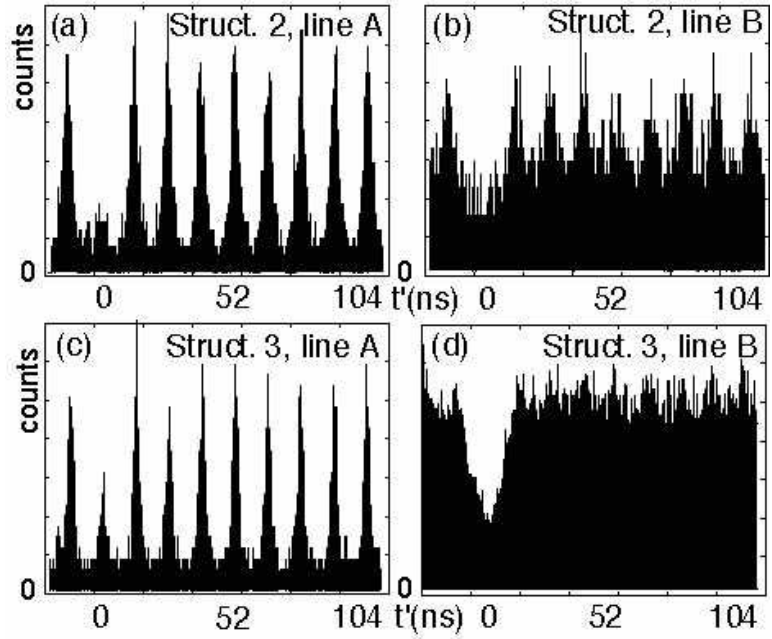}
\caption{Comparison of autocorrelation measurements for resonant dots (left) and off-resonant dots in PC (right).  (a,b) Structure 2, coupled and decoupled QDs. (c,d) Structure 3, coupled and decoupled QDs.  In all measurements, the pump was at 750\unit{nm} (above GaAs bandgap) and repeated at 13\unit{ns} intervals.  Background was not subtracted.} 
\label{fig:QD_cav_g2}
\end{figure}

To verify that the observed emission results from single emitters, we measured photon statistics by the autocorrelation function $g^{(2)}(t')=\braket{I(t) I(t+t')}/\braket{I(t)}^{2}$.  For short time scales, this function is measured from the coincidence rate between the two detectors of a Hanbury-Brown and Twiss interferometer.  A start-stop scheme measures time delay $t'=t_1-t_2$ between detection events.  The laser repetition period is 13\unit{ns} (Fig.\ref{fig:PL_setup}).  Fig. \ref{fig:QD_cav_g2}(a) presents the coincidence histogram for the cavity-coupled emission line A.  The antibunching of $g^{(2)}_A(0)\approx 0.14 <\fr{1}{2}$ at zero delay time indicates that the emission is indeed from a single emitter.  Lines B (Fig. \ref{fig:QD_cav_g2}(a)) and C  \cite{supplemental_section} are also single emitters with $g^{(2)}_B(0) \sim 0.04$ and  $g^{(2)}_C(0) \sim 0.03$ (estimated by exponential fits to the autocorrelation data).  


This pattern of short-lived coupled, and long-lived decoupled excitonic lines was observed in many other structures.  For instance, for structure 3,  the PL spectrum (Fig.\ref{fig:PL}(c)) shows a coupled single exciton line A, while a second line B at 932\unit{nm} is clearly decoupled.  Line A again has short lifetime $\tau_A =1.70$\unit{ns}, while B has very long $\tau_B =7.96$\unit{ns} (Fig. \ref{fig:QD_cav_g2}(c,d)).  Again, single photon behavior is apparent, with $g^{(2)}_A(0) = 0.23$ and $g^{(2)}_B(0) = 0.05$.   As an example of shortened lifetime, we show line A of Struct. 4.  The low pump-intensity spectrum of Fig.\ref{fig:PL}(d) shows coupling to the x-dipole cavity.  To increase spectral coupling, we shifted line A from 884.67 to 885.54\unit{nm} by temperature-tuning from 6 to 40K \cite{Vuckovic2003b}.  Lifetime is sharply reduced to 210\unit{ps} (Fig.\ref{fig:lifetime}(b)), roughly eight times shorter than in bulk GaAs, and photon coincidence is antibunched to $g^{(2)}(0) \approx 0.16$.

\begin{figure}[htbp]
    \includegraphics[width=4in]{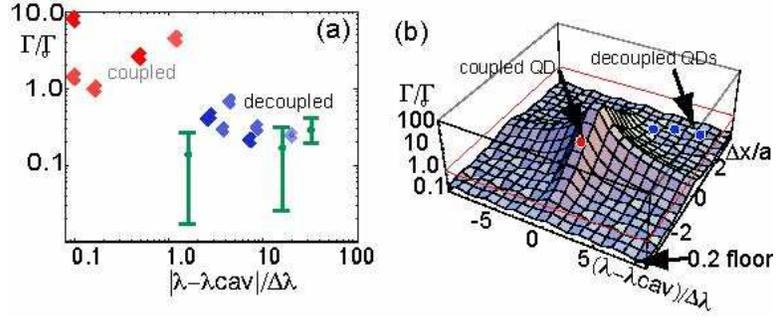}
    \caption{SE rate modification.  (a) Experimental (circles) and calculated (bars) data of $\Gamma/\Gamma_0$ of single QD exciton lines vs. spectral detuning (normalized by the cavity linewidth $\Delta \lambda$).  Coupling was verified by spectral alignment and polarization matching.  Data for shaded points are presented in \cite{supplemental_section}.  (b) Illustration of the predicted SE rate modification in PC as function of normalized spatial and spectral misalignment from the cavity ($a$ is the lattice periodicity).  This plot assumes $Q=1000$ and polarization matching between the emitter dipole and cavity field.  The actual SE rate modification varies significantly with exact QD location.}
    \label{fig:Enhancement}
\end{figure}

Off-resonant dots see up to five-fold lifetime enhancement.  To explain these increased lifetimes, we again turn to Eq.\ref{eq:purcell_factor}.  The first term for emission into the cavity mode vanishes as the QD exciton line is far detuned from $\lambda_{cav}$.  The second term $F_{PC}$ is reduced below unity due to the diminished LDOS in the PC for emission inside the PC bandgap, leading to longer lifetime.  This SE rate modification is illustrated in Fig.\ref{fig:Enhancement}(b).  

We verified the lifetime modification theoretically by FDTD analysis of a classical dipole in the PC with the x-dipole cavity.  The simulation is based on the result that the quantum electrodynamical and classical treatments of SE emission yield proportional results, so that the SE rate is related to the classical dipole radiation power by $\Gamma_{SE}^{PC}/\Gamma_{SE}^{bulk} = P_{classical}^{PC}/P_{classical}^{bulk}$ for bulk GaAs and the PC\cite{Xu1998}.  In these simulations, we replicated the conditions under which the experimental data were obtained: 200 dipoles were placed at random positions and orientations in a photonic crystal structure, roughly covering the focal size over which we collected in the experiment.  The same type of single-defect cavity with low $Q\approx320$ was at the center.  With the dipoles radiating at frequency $\fr{2\pi c}{\lambda}$ and detuned from from the cavity, we simulated the averaged emitted power and were thus able to calculate the spatial average over all dipoles (QDs) of $\Gamma/\Gamma_0=\Gamma_{SE}^{PC}/\Gamma_{SE}^{bulk}$.  To find the variance, we repeated this simulation for single emitters at a range of locations, one at a time.  For decoupled QDs, we found average lifetime suppression to $\Gamma/\Gamma_0 \sim 0.20$.   These results, plotted in Fig. \ref{fig:Enhancement}(a), agree well with our experimental observations and confirm earlier theoretical predictions of SE lifetime suppression inside the PC bandgap of a similar structure\cite{Lee2000}.  The TE bandgap of our structures extends from $\sim$784\unit{nm} to $\sim$1045\unit{nm}, calculated by FDTD simulations.  Outside the bandgap, we expect $\Gamma/\Gamma_0$ to return closer to the bulk value, though we cannot observe this effect as the QD distribution does not reach outside the bandgap. 

Due to spatial misalignment between the QD and cavity, most of the resonant dots presented here have moderate emission rate enhancement.  This misalignment is estimated at about one lattice spacing based on the cavity mode pattern (Fig.\ref{fig:FDTD_cav}).  Considering the low probability of QD-cavity coupling, these modest rate enhancements are not surprising.  With better spatial and spectral matching, Eq.\ref{eq:purcell_factor} predicts that a rate enhancement of up to $\sim$ 400 is possible.  Beyond that, Eq.\ref{eq:purcell_factor} is no longer valid as the system approaches the strong coupling regime near $\kappa \sim |g_{cav}(\vec{r}_A)|$.



In conclusion, we show that by designing a suitable photonic crystal environment, we can significantly modify SE rate of embedded quantum dots.  For QDs coupled to a PC defect cavity, we observe up to 8 times faster SE rate and demonstrate antibunching.  This coupled system promises to increase out-coupling efficiency and photon-indistinguishability of single photon sources.  In contrast, for individual off-resonant QDs in the PC, we show up to five-fold lifetime enhancement as a result of the diminished LDOS in the photonic bandgap.  This extended lifetime may find applications in QD-based photonic devices (e.g., switching), including quantum information processing (e.g., quantum memory).  It also shows that reported QD lifetime reduction due to surface proximity effects\cite{Wang2004} should not limit the performance of PC-QD single photon sources.  We find good agreement between these results and FDTD simulations of SE in the photonic crystal.  

Financial support was provided by the MURI Center for photonic quantum information systems (ARO/ARDA Program DAAD19-03-1-0199), JST, SORST project on quantum entanglement, NTT Basic Research Laboratories, NSF grants ECS-0424080 and ECS-0421483, as well as NDSEG and DCI fellowships.  Sample growth was supported by the IT Program, MEXT.

\bibliographystyle{unsrt}
\bibliography{references.bib}

\clearpage

%

\end{document}